%% file: Letter-m12-221121.tex
\documentclass[letterpaper]{article}
\pdfoutput=1
\usepackage{dcolumn}
\usepackage{bm}
\usepackage{caption}

\usepackage{graphicx}
\usepackage{amssymb,amsmath}
\usepackage{multirow}
\usepackage{multicol}
\usepackage{nonfloat}
\usepackage{cite,color,url}
\usepackage[colorlinks=true
,urlcolor=blue
,anchorcolor=blue
,citecolor=blue
,filecolor=blue
,linkcolor=blue
,menucolor=blue
,linktocpage=true
,pdfproducer=medialab
,pdfa=true
]{hyperref}

\usepackage{epsfig,psfrag,rotating,soul}
\usepackage{rotfloat}

\evensidemargin \oddsidemargin
\allowdisplaybreaks

\input{paperdef}

\psfrag{lk}{$l_k$}
\psfrag{lm}{$l_m$}
\psfrag{blk}{$\bar {l}_k$}
\psfrag{blm}{$\bar {l}_m$}

\def\thefootnote{\fnsymbol{footnote}}

\let\OLDthebibliography\thebibliography
\renewcommand\thebibliography[1]{
  \OLDthebibliography{#1}
  \setlength{\parskip}{0pt}
  \setlength{\itemsep}{0pt plus 0.3ex}
}

\newcommand\myfigure[1]{%
\medskip\noindent\begin{minipage}{\columnwidth}
\centering%
#1%
\end{minipage}\medskip}

\begin{document}

\begin{flushright}
IFT-UAM/CSIC-22-065\\
\end{flushright}

\vspace{0.5cm}

\begin{center}

\begin{Large}
  \textbf{\textsc{Sensitivity and constraints to the 2HDM soft-breaking\\[.3em]
      \boldmath{$Z_2$ parameter $m_{12}$}}}
\end{Large}

\vspace{1cm}

{\sc
F. Arco$^{1,2}$%
\footnote{\tt \href{mailto:francisco.arco@uam.es}{francisco.arco@uam.es}}%
, S. Heinemeyer$^{2}$%
\footnote{\tt \href{mailto:sven.heinemeyer@cern.ch}{Sven.Heinemeyer@cern.ch}}%
, M.J. Herrero$^{1,2}$%
\footnote{\tt \href{mailto:maria.herrero@uam.es}{maria.herrero@uam.es}}%
}

\vspace*{.7cm}

{\sl
$^1$ Departamento de F\'{\i}sica Te\'orica and Instituto de F\'{\i}sica Te\'orica,\\
Universidad Aut\'onoma de Madrid, Cantoblanco, 28049 Madrid, Spain

\vspace*{0.1cm}

$^2$ Instituto de F\'{\i}sica Te\'orica, IFT-UAM/CSIC,\\
Universidad Aut\'onoma de Madrid, Cantoblanco, 28049 Madrid, Spain

}

\end{center}

\vspace*{0.1cm}

\begin{abstract}
\noindent
In this letter we study the specific sensitivity and constraints to the
soft breaking $Z_2$ parameter $m_{12}$ of the Two Higgs Doublet
Model (2HDM). 
$m_{12}$ is considered here as an input parameter of the model together
with the masses of the Higgs bosons,  $m_h$, $m_H$, $m_A$, $m_{H^\pm}$, 
the ratio of the two Higgs vacuum expectation values, $\tan \beta$, 
and $\CBA$,  with $\al$ and $\be$  the mixing angles of the $\CP$-even
and $\CP$-odd 
Higgs sector,  respectively. 
We explore in particular the sensitivity to $m_{12}$ in the decays of
the lightest Higgs boson $h$ to two photons,  assuming: 1) that 
$h$ is the state observed at the LHC at $\sim 125 \gev$, 
and 2) the $h$ couplings to the SM particles are SM-like,
by going to the alignment limit, $\CBA = 0$.
The aim of this work is to demonstrate possible
constraints on $m_{12}$ from the present
measurement of the di-photon signal strength,
$\mu_{\gamma \gamma}$. These constraints are relevant in the region
of the 2HDM parameter space that is allowed by all the other
constraints,  specifically,  
theoretical constraints as well as experimental constraints from
  LHC Higgs rate measurements, Higgs boson searches,
flavor physics and precision observables.  
The exploration is performed in the four different Yukawa
types of 2HDM,  where the
allowed region by all the other constraints depends on the model type. 
We also analyze the case that the Higgs-boson mass spectrum
  accommodates a possible new world average of $m_W$ including the recent
  CDF measurement.
\end{abstract}

\vspace*{0.5cm}

\def\thefootnote{\arabic{footnote}}
\setcounter{footnote}{0}

\begin{multicols}{2}

\section{Introduction}
\label{sec:intro}

The Two Higgs Doublet Model (2HDM) is one of the most studied
extensions of the Standard Model (SM) of particle physics,
and it has been explored exhaustively for many years (see, 
e.g.,~\cite{Gunion:1989we,Aoki:2009ha,Branco:2011iw} for reviews). This model
contains two Higgs doublets,  $\Phi_1$ and $\Phi_2$,  that accommodate
the five physical Higgs bosons: the  
two $\CP$-even $h$ and $H$, the $\CP$-odd $A$, and the pair of
charged Higgs bosons, $H^\pm$.  The lightest Higgs boson   $h$ is
usually  assumed to be SM-like 
 with a mass set to the experimentally measured value of $\sim 125 \gev$
 (see, e.g.,\cite{PDG2022}, for a recent summary).  Other parameters are
the two mixing angles $\al$ and $\be$ which diagonalize the $\CP$-even
and $\CP$-odd 
Higgs boson sector, respectively, and 
$\tb=v_2/v_1$,  the ratio of the two Higgs vacuum expectation
values,  which are related to the SM vev by $v=\sqrt{(v_1^2+v_2^2)}$.  

In order to avoid flavor changing neutral currents (FCNC) at the
tree-level, a $Z_2$~symmetry is imposed~\cite{Glashow:1976nt}.
When this symmetry is allowed to be softly broken in the Higgs
potential,  an extra 2HDM parameter $m_{12}$ with dimension of mass is
included in the Higgs potential by a new term
$\sim \msq (\Phi_1^\dagger \Phi_2 + \Phi_2^\dagger \Phi_1)$.
The relevance of this $m_{12}$ parameter is clear,  since it
provides the mixing between the two Higgs doublets which in
turn can give rise to FCNC beyond the tree level. 
Therefore limiting
the size of this $m_{12}$ parameter is a very important issue,
complementing constraining  the physical Higgs  boson masses. 

Four types of the 2HDM can then be realized, 
depending on how this $Z_2$~symmetry is implemented into the fermion sector:
types~I-IV~\cite{Aoki:2009ha}. 
Once $m_h$ and $v$ are set to $m_h \simeq 125 \gev$ and $v\simeq 246
\gev$,  the four 2HDM types can be fully described in terms of six input
parameters which we choose here to be: $\cos(\beta-\alpha)$,
$\tan\beta$,  $m_H$,  $m_A$, $m_{H^\pm}$ and $m_{12}$.  

So far, the experimental searches for these new Higgs bosons have not
found any indication of their existence,  neither direct nor indirect,
setting limits
on the 2HDM parameters.   There are many analysis in
the literature of these constraints,  including various sets of data,
various choices of benchmark planes,  and various ways to implement the
theoretical constraints on these parameters (considering,  for instance,
a tree-level or one-loop analysis). 
We will focus here on the analysis
of the 2HDM constraints presented in \citere{Arco:2020ucn},
which has been recently updated to the four types of 2HDM in 
\citere{Arco:2022}. This analysis was performed
with the goal of deriving the allowed ranges of the triple Higgs
boson coupling. These couplings can have interesting consequences for the
double Higgs production at $e^+e^-$ colliders,  as it was shown in
\citere{Arco:2021bvf}.  Notice that with our choice of input parameters,
and using the physical basis,  these triple Higgs couplings
$\lambda_{h_ih_jh_k}$ are derived quantities,   and 
it was demonstrated in \citeres{Arco:2020ucn,Arco:2022} that some of
them can reach sizeable values.  In particular,  it was shown in
\citere{Arco:2022} that the coupling of the lightest Higgs to the
charged Higgs bosons $\lambda_{hH^+H^-}$ can be as large as $\sim
20-30$,  in the four 2HDM types  and yet  be compatible with all the
present constraints. 

Our main motivation in this letter is to perform a dedicated study on
the sensitivity to the soft-breaking $Z_2$ parameter $m_{12}$ in order to
analyze the constraints on this parameter.  In this
analysis we will include all available theoretical and
experimental constraints on equal footing for
the four 2HDM Yukawa types,  following the same procedure as in
\citere{Arco:2022}.   The most important  difference with respect to our
previous analysis  is that here, in addition to the full set of
constraints specified in \citere{Arco:2022},  we will select and analyze
in full detail one particular observable which carries the highest
sensitivity to the $m_{12}$ parameter.  This observable is the decay
width of the lightest Higgs boson to two photons,
$\Gamma(h \to \gamma \gamma)$
that proceeds in the 2HDM at the one-loop level. The prediction
of this partial width within the SM is known since long
ago~\cite{Ellis:1975ap}.  Additional contributions from charged  
scalars are also known since long ago~\cite{Shifman:1979eb}
(see also \citere{Gunion:1989we} for a review).  The
sensitivity to $m_{12}$ occurs via the charged Higgs boson loop and
precisely through the triple Higgs coupling $\lambda_{hH^+H^-}$.  Our
goal here is to determine and update the sensitivity to $m_{12}$ via
the prediction of the ratio  
BR$(h \to \gamma \gamma)$ in the 2HDM,  and conclude on the corresponding
constraints to $m_{12}$ from the comparison with the present
experimental signal strength $\mu_{\gamma \gamma}$~\cite{PDG2022}. 
Another important difference in the present analysis with
respect to \citere{Arco:2022} is that we restrict ourselves here to
the alignment limit~\cite{Bernon:2015qea} where
$\cos(\beta-\alpha) \to 0$.  This limit is in very good agreement
with the present data from LHC rate measurements,  that restrict the
size of $\cos(\beta-\alpha)$ 
to values lower than ${\cal O}(10^{-2})$  for type~II and~III
and lower than ${\cal O}(10^{-1})$  for type~I and~IV, 
see for instance \citeres{Arco:2020ucn,Arco:2022}.
In the alignment limit the $h$
couplings to SM particles approach the corresponding SM couplings.
However,
there are important  implications for beyond SM physics from the triple
Higgs couplings of $h$  to the heavy Higgs bosons,  and here in
particular from
$\lambda_{hH^+H^-}$,  which carries the sensitivity to
$m_{12}$ via the $h \to \gamma \gamma$ decay.  


\section{The \boldmath{$m_{12}$} soft-breaking parameter in the 2HDM}
\label{sec:2hdm}

The $m_{12}$  soft-breaking parameter appears in the Higgs potential of
the 2HDM.  
This potential for the  $\cp$ conserving 2HDM is 
given in terms of eight parameters by~\cite{Branco:2011iw}:
\begin{align}
&V = m_{11}^2 (\Phi_1^\dagger\Phi_1) + m_{22}^2 (\Phi_2^\dagger\Phi_2)-\msq (\Phi_1^\dagger
\Phi_2 + \Phi_2^\dagger\Phi_1) \nonumber \\
&+ \frac{\la_1}{2} (\Phi_1^\dagger \Phi_1)^2 +
\frac{\la_2}{2} (\Phi_2^\dagger \Phi_2)^2  +  \la_3
(\Phi_1^\dagger \Phi_1) (\Phi_2^\dagger \Phi_2)  \nonumber \\
&+  \la_4
(\Phi_1^\dagger \Phi_2) (\Phi_2^\dagger \Phi_1) + \frac{\la_5}{2}
[(\Phi_1^\dagger \Phi_2)^2 +(\Phi_2^\dagger \Phi_1)^2]\,,  
\label{eq:scalarpot}
\end{align}
\noindent
where $\Phi_1$ and $\Phi_2$ denote the two $SU(2)_L$ scalar doublets.
As mentioned above, the occurrence of tree-level flavor
changing neutral currents (FCNC) is avoided by imposing a $Z_2$ symmetry 
on the scalar potential. 
The scalar fields transform as  $ \Phi_1 \to \Phi_1$,  $\Phi_2 \to - \Phi_2$.
The $Z_2$ symmetry, however, is softly broken by the $\msq$ term in
the Lagrangian. The extension of the $Z_2$ symmetry to the Yukawa
sector avoids tree-level FCNCs. 

Depending on the $Z_2$ parities of the fermions, this results in four
variants of 2HDM: type I, type II, 
type III (also known as type Y or flipped) and 
type IV (also known as type X or lepton-specific)~\cite{Aoki:2009ha}, 
that are summarized in \refta{tab:types}.\\

\noindent
\begin{minipage}{\columnwidth}
\begin{center}
\begin{tabular}{lccc} 
\hline
  & $u$-type & $d$-type & leptons \\
\hline
Type~I 									& $\Phi_2$ & $\Phi_2$ & $\Phi_2$ \\
Type~II 								& $\Phi_2$ & $\Phi_1$ & $\Phi_1$ \\
Type~III								& $\Phi_2$ & $\Phi_1$ & $\Phi_2$ \\
Type~IV								& $\Phi_2$ & $\Phi_2$ & $\Phi_1$ \\
\hline
\end{tabular}
\captionof{table}{Allowed fermion couplings in the four 2HDM types.}
\label{tab:types}
\end{center}
\end{minipage}
\\

After the consideration of the minimization conditions of the above
potential for the implementation of the electroweak symmetry breaking
(with $v \simeq 246 \gev$),  
using the physical basis for the Higgs sector
and  identifying the lightest $\cp$-even Higgs boson,
$h$,  with the one observed at $\sim 125 \gev$,   the free
parameters of the 2HDM are reduced to the following six:
\begin{equation}
c_{\be-\al} \; , \; \tb \;,
 \; \MH \;, \; \MA \;, \; \MHp \;, \; m_{12} \;.
\label{eq:inputs}
\end{equation}
Here and from now on we use the short-hand notation
$s_x = \sin(x)$, $c_x = \cos(x)$.

The triple Higgs couplings in the physical basis,  $\la_{h_i h_j h_k}$,
are then derived quantities that can be evaluated from the
input parameters in \refeq{eq:inputs}.  We employ here the tree level
predictions for these 
triple couplings whose specific expressions and notation are taken from
\citeres{Arco:2020ucn,Arco:2022}.   In particular, the triple coupling of the
lightest Higgs boson to the charged Higgs bosons is given by: 
\begin{align}
  \lambda_{hH^+H^-} &= \frac{1}{v^2} \big\{\left(m_h^2+2 m_{H^\pm }^2-2
                     \bar{m}^2\right)s_{\be -\al }  \nonumber \\
	&+ 2 \cot 2 \be \left(m_h^2-\bar{m}^2\right) c_{\be -\al }\big\}\,,
\end{align}
where the soft-breaking parameter $m_{12}$
enters via $\bar{m}$,  which is defined by:
\begin{align} 
 \bar{m}^2 &= \frac{\msq}{\sin\be\cos\be} \,.
\label{eq:mbar}
\end{align} 
The introduction of this auxiliary parameter $\bar{m}$ in the present
analysis is convenient for practical reasons, and because it summarizes
the effect of $m_{12}$ and $\tan \beta$ in $\lambda_{hH^+H^-}$.   
Finally,  when assuming the alignment limit,  the triple coupling
$\lambda_{hH^+H^-}$ simplifies to: 
\begin{equation} 
  \lambda_{hH^+H^-}^{\rm align} =
  \frac{1}{v^2} \left(m_h^2+2 m_{H^\pm}^2-2 \bar{m}^2\right). 
\label{eq:lalign}
\end{equation}
Therefore,  the size of this relevant triple coupling is directly
related to the splitting ($m_{H^\pm}^2-{\bar m}^2$).


\section{\boldmath{$h \to \gamma \gamma$ } in the alignment\\
  limit: sensitivity to \boldmath{$m_{12}$}} 
\label{htogaga-alignment}

As it is well known,  the di-photon decay of the neutral Higgs boson
$h$ proceeds at the one-loop level with contributions from
fermions, $W$~bosons, and charged Higgs bosons.  The partial
width $\Gamma(h \to \gamma \gamma)$
was computed long ago~\cite{Ellis:1975ap}, and it is 
usually written in terms of the corresponding form factors for the
fermion,  $A^f$,  $W$ boson, $A^{W^\pm}$, and charged Higgs boson loop
contributions, $A^{H^\pm}$.  Generically (see, for instance, the full
formula in \cite{Gunion:1989we,Eriksson:2009ws}) it can be written as: 
\begin{equation}
\Gamma(h \to \gamma \gamma)= \frac{\alpha^2\, m_h^3}{256\, \pi^3\, v^2}
\left| A^f+A^{W^\pm}+ A^{H^\pm}\right|^2. 
\end{equation}

The prediction  for this partial width in the
2HDM is in general different from the SM Higgs di-photon width.
However, in
the alignment limit,  the only difference between the 2HDM and the SM
prediction comes from the contribution of the charged Higgs boson loops,
where the corresponding contribution can be written as:  
\begin{equation}
A^{H^\pm}=-\lambda_{hH^+H^-}^{\rm align} \frac{v^2}{2m_{H^\pm}^2} F_0(\tau_{H^\pm})\, ,
\end{equation}
where,  $\tau_{H^\pm}=m_h^2/(4m_{H^\pm}^2)$,  and 
\begin{equation}
F_0(\tau)=\tau^{-1}\left(\tau^{-1} {\rm arcsin}^2 (\sqrt{\tau})-1 \right)\,.
\end{equation}
Therefore, the relevant parameters that will give the main differences 
between the 2HDM prediction of $\Gamma(h \to \gamma \gamma)$
in the alignment limit  with respect to the SM prediction are clearly,
$m_{H^\pm}$,  $m_{12}$ and $\tan\beta$.  The sensitivity to $m_{12}$ and
$\tan\beta$ in the partial width comes jointly from the $\bar{m}$ that
appears in $ \lambda_{hH^+H^-}$ which, as discussed above,
is a derived parameter
\footnote{The sensitivity to the $\lambda_5$ coupling of the potential
(which is related to $\bar m$) in $h\to\gamma \gamma$ decays and in the 
alignment limit was also studied previously in \cite{Arhrib:2003vip} 
reaching similar conclusions on the effects of this parameter on the 
$\br(h \to \gamma \gamma)$.  }.
Notice that the other 2HDM mass parameters,
$m_H$ and $m_A$, do not enter at leading order in the partial width
$\Gamma(h \to \gamma \gamma)$, or in the total $h$~width as long as the
alignment limit is imposed, no EW higher order corrections are included,
and the decay channel $h\to AA$ is not kinematically open.   
In our analysis for the numerical evaluations of the
BR$(h \to \gamma \gamma)$ we use the code
\texttt{2HDMC-1.8.0}~\cite{Eriksson:2009ws}.   

The main aspects of the sensitivity to $m_{12}$ in BR$(h \to \gamma
\gamma)$ are summarized in \reffi{BRtogaga}. 
In the upper row we show the $m_{12}$-$\MHp$ plane (left) and the
$m_{12}$-$\tb$ plane (right) for $\MH = \MA = \MHp$ with $\tb = 3$
(left) and $\MHp = 550 \gev$ (right). The color coding indicates
$\br(h \to \ga\ga)$.  The blue line indicates the ``SM limit'',  which 
we define here by setting the input $m_{12}$ such that $\lahHpHm = 0$.
One can observe that for
fixed $m_{H^\pm}>200$~GeV  and  $\tan \beta >1$,  the BR grows
with $m_{12}$ and gets values clearly above the SM value in a large
region of the 2HDM parameter space.
 
In the middle row we show 
$\br(h \to \ga\ga)$ as a function of $\MHp$
for $\tb = 1$ (left) and $\tb = 50$ (right)
for various choices of $m_{12}$, where the ``SM
limit'' is represented by the horizontal blue lines.  
One can observe that for fixed $m_{12}$ (solid lines)
the coincidence of the 2HDM prediction with the SM value does not occur
in the very heavy charged Higgs-boson limit
where $m_{H^{\pm}}  \to \infty$,
since $\lambda_{hH^+H-}$ being a derived quantity grows with
$m_{H^\pm}$,
(and at some point becomes non-perturbative),
and as a consequence the  BR$(h \to \gamma \gamma)$ tends
to a constant value different than the SM value.  In this heavy
charged Higgs-boson limit
the loops with charged Higgs bosons are not really decoupled.   
In fact, $A^{H^\pm} \to -1/3$ in the limit $m_{H^\pm}\to\infty$.
On the other hand, in a different setting of the input parameters with
$\MHp = \bar m$ (dashed olive 
lines) the triple Higgs coupling $\lambda_{hH^+H^-}^{\rm align}$
takes a fixed
value of $m_h^2/v^2$ (see \refeq{eq:lalign}) and,  for $\MHp \to \infty$,  the 
$\br(h \to \ga\ga)$ tends to the same constant value as the above
defined ``SM limit''.   
Therefore,  in summary,  to quantify the sensitivity to $m_{12}$ in
BR$(h \to \gamma \gamma)$ the optimal comparison is to estimate the departure of
the 2HDM prediction with respect to  the ``SM limit'' defined in this section.
The concrete value that we obtain for $\br(h\to\ga\ga)$ in this is ``SM limit'' 
is $2.181\times 10^{-3}$. 
 \end{multicols}
\myfigure{
\hspace{6mm}
\includegraphics[scale=0.54]{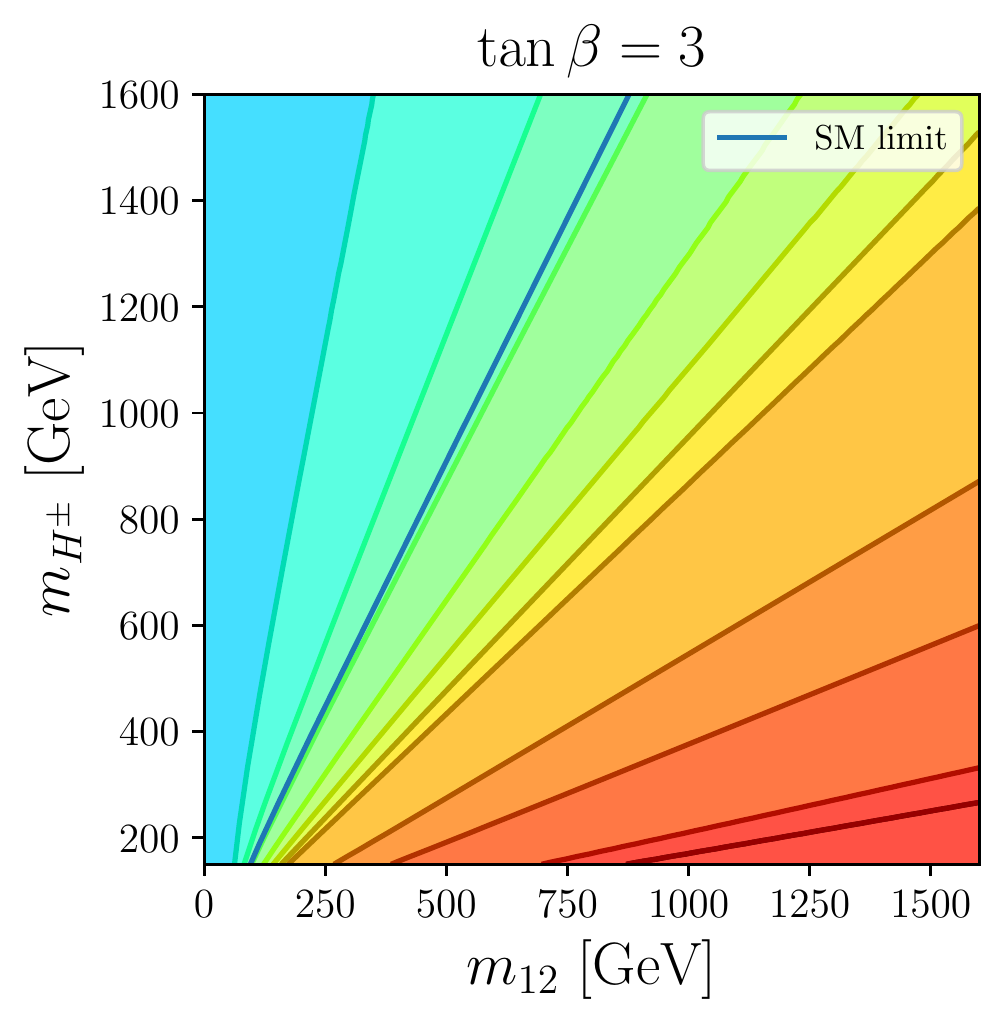}
\includegraphics[scale=0.54]{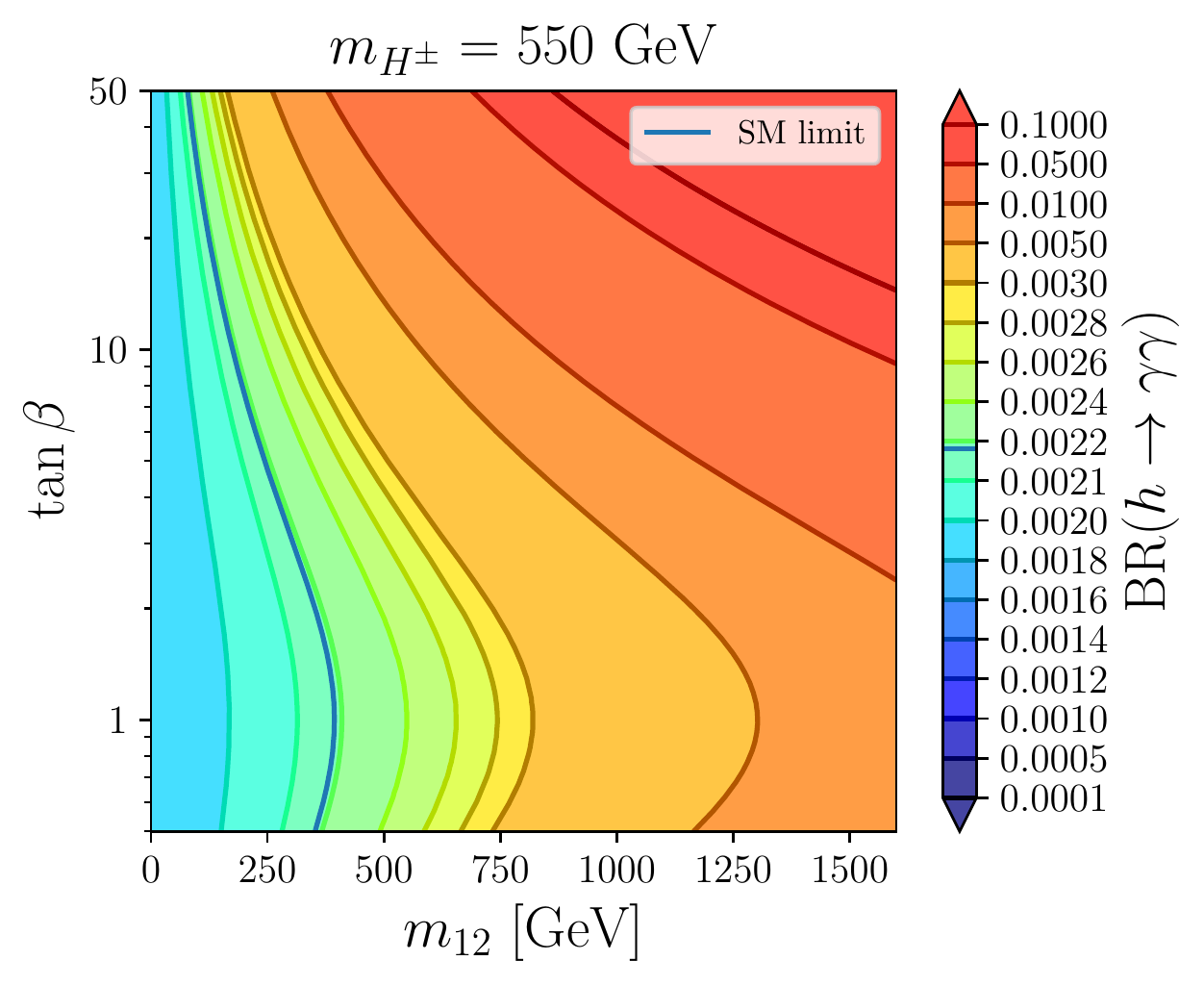}
\includegraphics[scale=0.52]{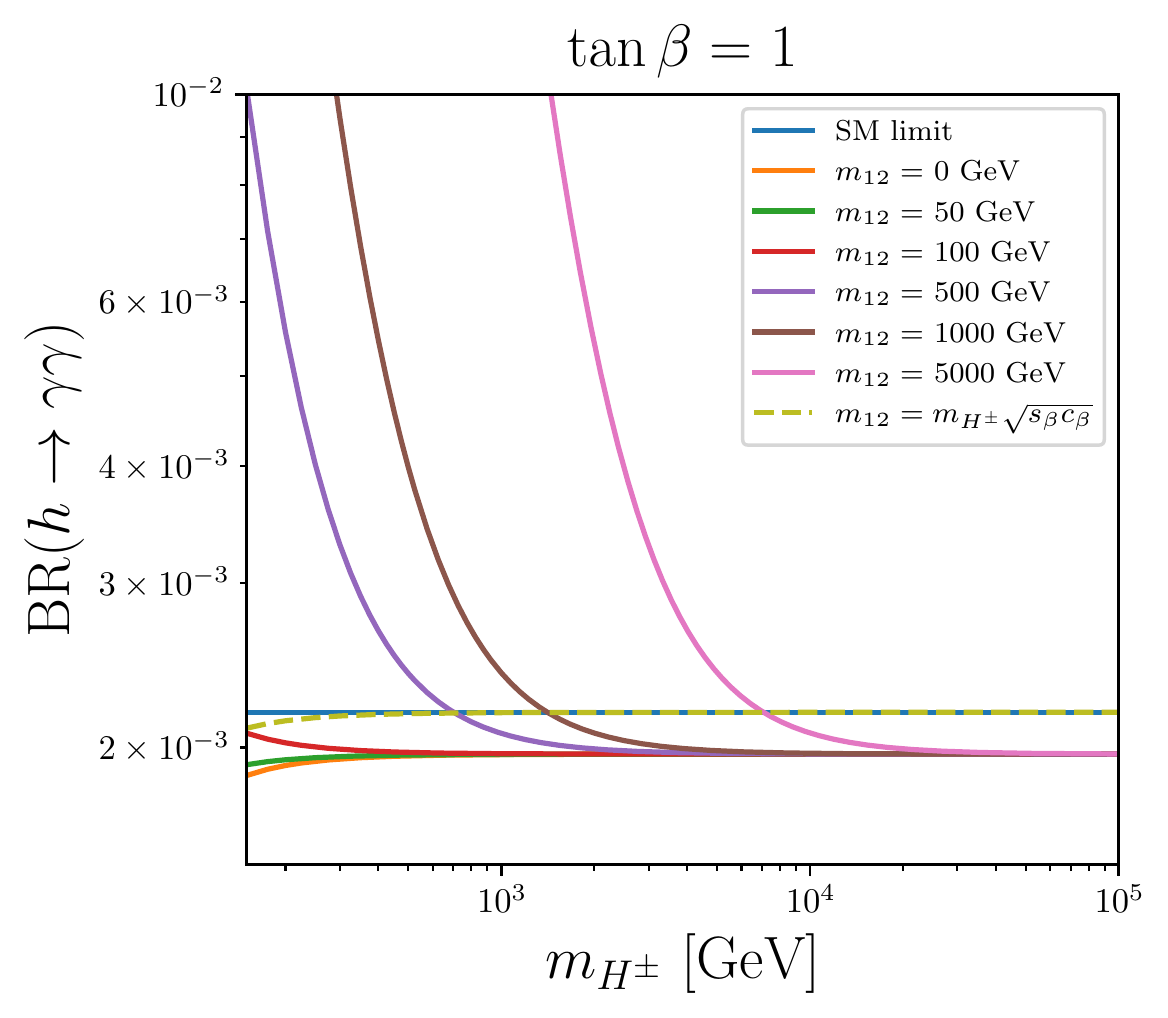}
\includegraphics[scale=0.52]{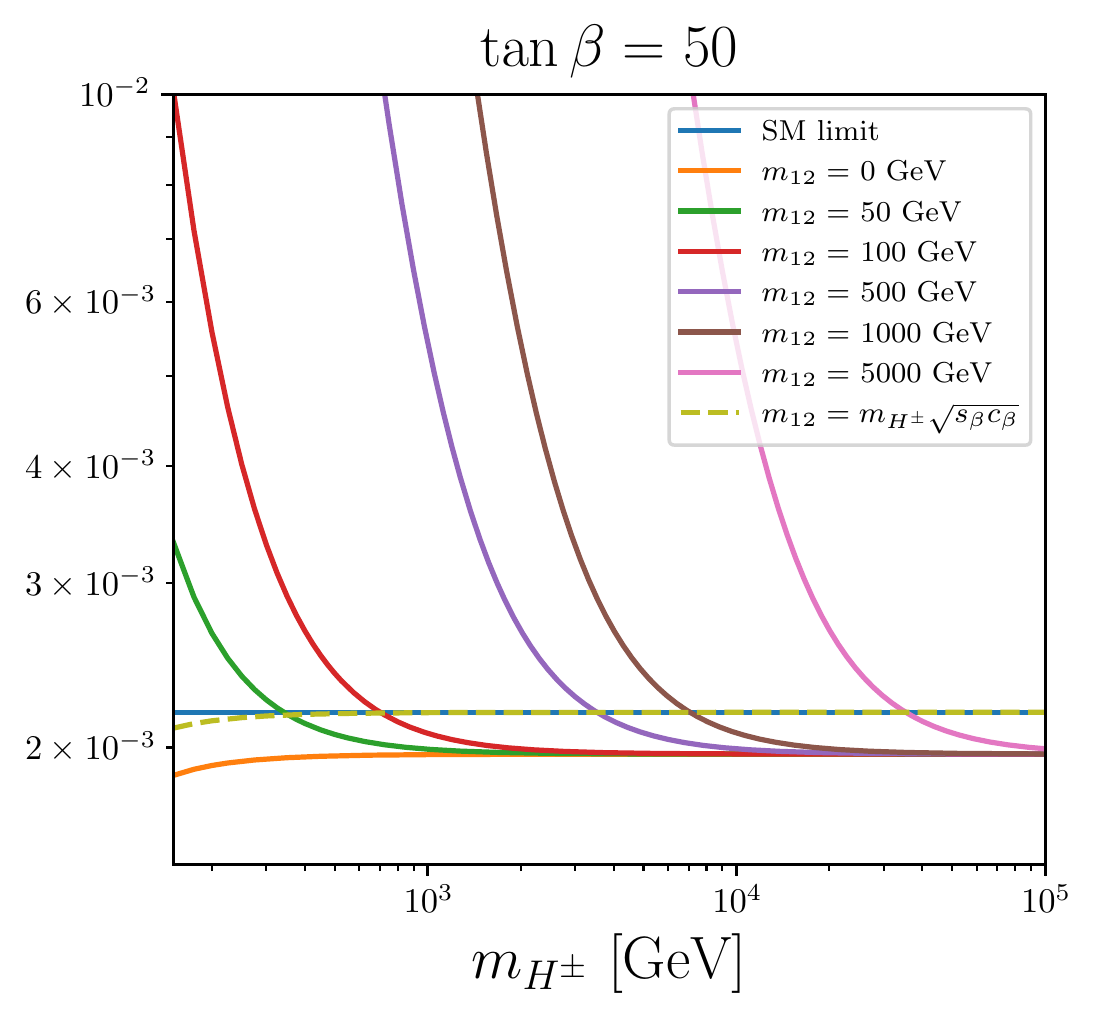}

\hspace{8mm}
\includegraphics[scale=0.54]{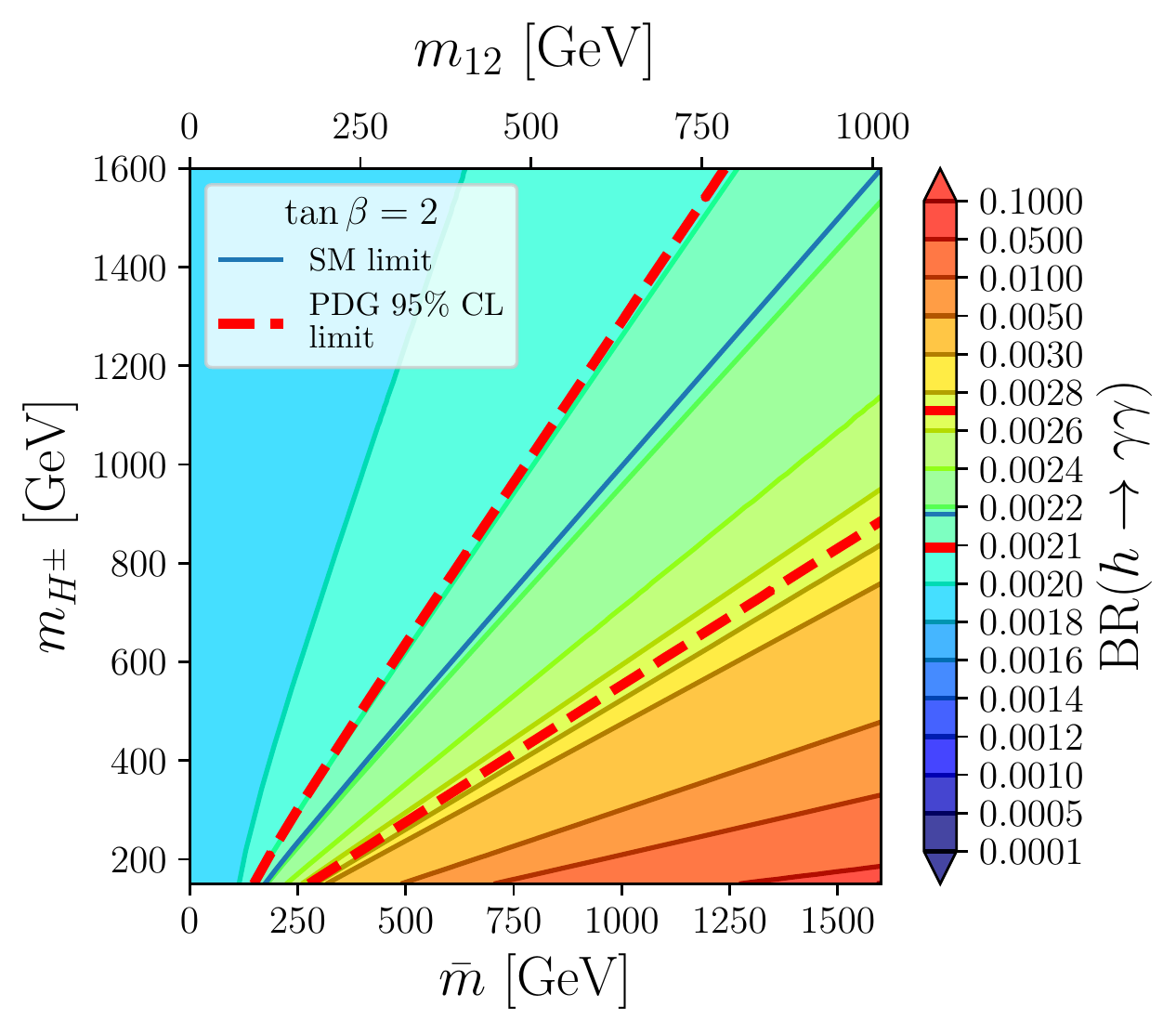}
\includegraphics[scale=0.54]{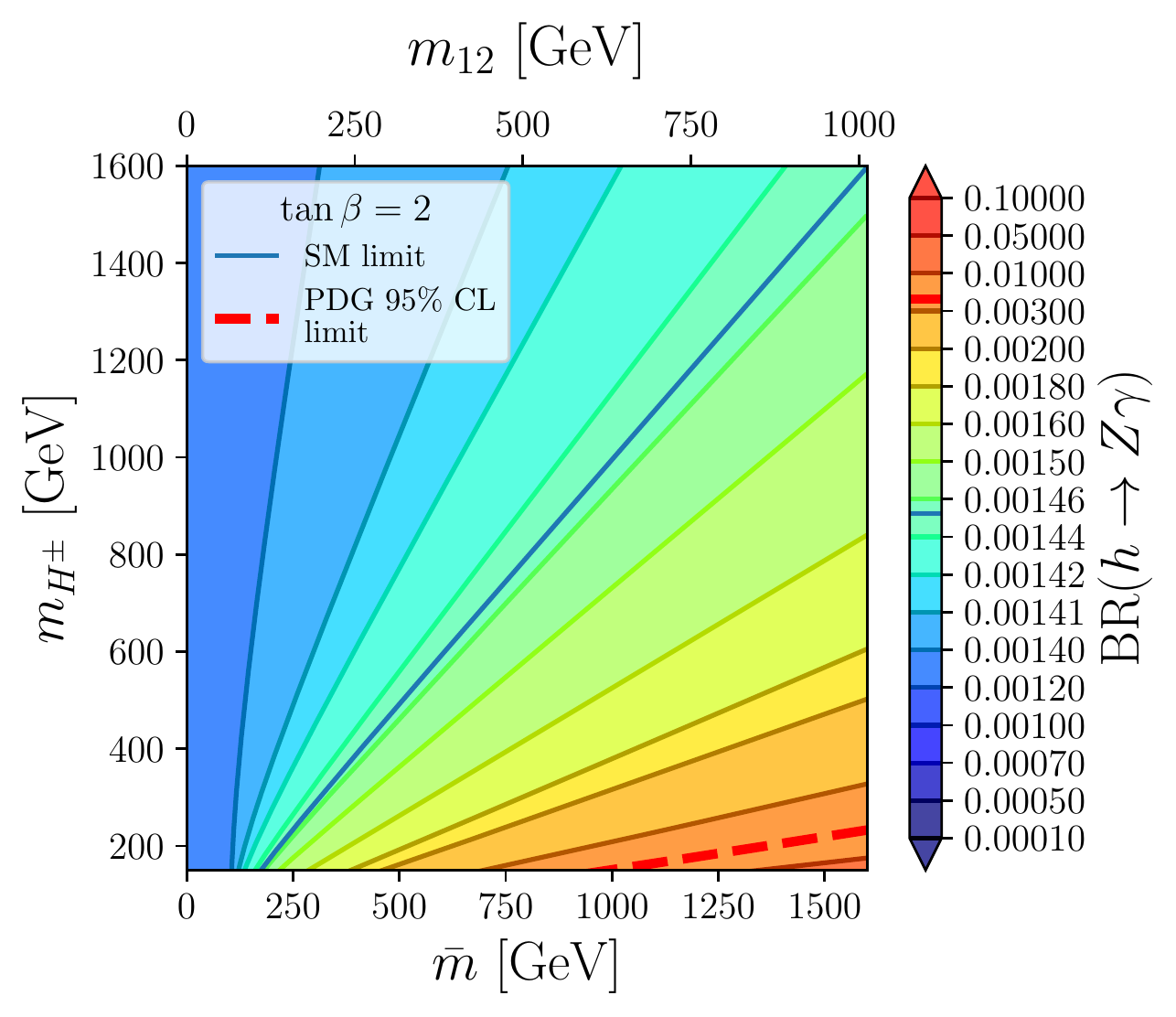}
\figcaption{BR$(h \to \gamma \gamma)$ in the 2HDM, in the alignment
  limit,  as a function of the 
  relevant parameters $m_{H^\pm}$,  $\tan\beta$ and $m_{12}$ (upper and middle
  rows).  In the lower row we compare the prediction for BR$(h \to
    \gamma \gamma)$ (left) to the prediction for BR$(h \to Z \gamma)$
    (right) in the ($\bar m$,$\MHp$) plane.}
\label{BRtogaga}
}
\begin{multicols}{2}
Finally,  in the lower left plot of \reffi{BRtogaga} the
prediction of BR$(h \to \gamma \gamma)$ is shown in the
($\bar m$-$\MHp$) plane, as before also in the alignment limit.
For comparison  in the lower right plot we
show the prediction for BR$(h \to Z \gamma)$ for the same choice of
parameters.  In the upper axes
of these two plots the corresponding $m_{12}$ values for $\tb =2$ are
displayed.  The SM limit contours indicated by the blue
lines where $\lahHpHm=0$ are included as well.  The dependence on
$\tb$ and $m_{12}$ in both cases enters through the parameter $\bar m$
present in $\lahHpHm$. Consequently,  as expected,  both
observables BR$(h \to \gamma \gamma)$ and BR$(h \to Z \gamma)$
display similar dependences on the relevant parameters $\MHp$, $\tb$
and $m_{12}$. The dashed red lines in these two plots
indicate the current 95\% CL limits that are extracted from the
corresponding signal strengths (which will be discussed in detail
in the next section).  In 
the $\gamma \gamma$ channel the allowed area lies inside the two
dashed red lines.  In the $\gamma Z$ channel the allowed area covers
practically the full ($\bar m$,$\MHp$) plane shown, except for very
low values of $m_{H^\pm} \lsim 200 \gev$ in the lower right
corner.  From these plots one can read off that in the allowed area for  
BR$(h \to \gamma \gamma)$ the predictions for BR($h\to Z\gamma$) give
values  in the interval  $1.43\times 10^{-3} - 1.58\times 10^{-3}$, which
are far from the present sensitivity.  This demonstrates that
BR$(h \to \gamma\gamma)$ is by far the most constraining observable
for $m_{12}$, and thus it will be the focus of this letter.

  
\section{Summary of the applied\\ constraints}
\label{constraints}

We summarize here the experimental and theoretical constraints on the
2HDM parameters  that are considered in the present analysis (more
details can be found in~\citeres{Arco:2020ucn, Arco:2022}).

\begin{itemize}
\item {\bf Constraints from electroweak precision data}\\
Constraints from the electroweak precision observables (EWPO)
can for ``pure'' Higgs-sector extensions of the SM, 
be expressed in terms of the oblique parameters $S$, $T$ and
$U$~\cite{Peskin:1990zt,Peskin:1991sw}.
Previous explorations of these parameters 
within the 2HDM were done long ago, see for instance \cite{Grimus:2007if,Grimus:2008nb}. 
In the 2HDM the $T$~parameter is the most
constraining parameter and requires either $\MHp \approx \MA$ or $\MHp
\approx \MH$. 
For the numerical evaluation of the $S$, $T$ and $U$ parameters in the
2HDM we use the code 
\texttt{2HDMC-1.8.0}~\cite{Eriksson:2009ws}
where the one-loop formulas are implemented.  These numerical
predictions for $S$, $T$ and $U$ are compared with the experimental
allowed values which we take from \cite{PDG2022}.
These constraints so far do not take into account the recent
measurement of $m_W$ by CDF~\cite{CDF2}.  However,  below we also analyze
the case of a possible new world average of $m_W$, where this
measurement is included.

\item {\bf Theoretical constraints}\\
Here the important constraints come from
tree-level perturbartive unitarity and stability of the vacuum.
They are ensured by an explicit test on the underlying Lagrangian
parameters, see \citere{Arco:2020ucn} for details.  For the evaluation
of these theoretical constraints we use our private code in which we
have implemented the formulas of the review \cite{Bhattacharyya:2015nca} 
(see also references therein, where the original computations can be
found),
following the procedure described in \citere{Arco:2020ucn}.   

\item {\bf Constraints from searches for BSM Higgs bosons}\\
The $95\%$ confidence level
exclusion limits of all relevant searches for BSM Higgs bosons
are included in the public code
\HB\,\texttt{v.5.9}~\cite{Bechtle:2008jh,Bechtle:2020pkv},
including Run~2 data from the LHC.
To test a (2HDM) parameter point, based on the masses, partial
  widths etc.\ of the Higgs bosons in the model, 
\HB\ (HB) determines which is the most
sensitive channel based on the expected experimental limits.  Only
this limit is then used to test whether a Higgs boson
is allowed or not at the $95\%$~CL.
The required input is calculated with the
help of \texttt{2HDMC}~\cite{Eriksson:2009ws}.

\item {\bf Constraints from the LHC Higgs rate measurements}\\
Any model beyond the SM has to accommodate the SM-like Higgs boson,
with mass and signal strengths as they were measured at the LHC.
In our scans the compatibility of the $\cp$-even scalar $h$ with a mass
of $125.09\gev$ with the measurements of LHC rate measurements
is checked with the code
\texttt{HiggsSignals v.2.6}~\cite{Bechtle:2013xfa,Bechtle:2020uwn} (HS). 
The code provides a
statistical $\chi^2$ analysis,
where the various measured LHC signal rates are compared with the
model prediction for the parameter point under consideration.
Again, the predictions of the 2HDM
have been obtained with {\tt{2HDMC}}~\cite{Eriksson:2009ws}.
As in \citere{Arco:2020ucn,Arco:2022}, we require that for a  parameter
point of the 2HDM to be allowed, the corresponding $\chi^2$ is within
$2\,\sig$ ($\De\chi^2 = 6.18$)
from the SM value.

\item {\bf Constraints from flavor physics}\\
Constraints from flavor physics have proven to be very significant
in the 2HDM mainly because of the presence of the charged Higgs boson.
Various flavor observables like rare $B$~decays, 
$B$~meson mixing parameters, $\br(B \to X_s \gamma)$, 
LEP constraints on $Z$ decay partial widths
etc., which are sensitive to charged Higgs boson exchange, provide
effective constraints on the available 
parameter space~\cite{Enomoto:2015wbn,Arbey:2017gmh}. 
Other observables like $\tau \to \mu \nu \nu$ have also been considered 
in the literature ~\cite{Enomoto:2015wbn,Abe:2015oca}.  However,  this $\tau$ decay 
is only relevant in type IV (lepton-specific) at very low values of 
$m_{H^\pm} < 200 \gev$, very low values of $m_A < 100 \gev$ 
and very large $\tan\beta$ above 40.
Here we take into account the decays $B \to X_s \gamma$ and
$B_s \to \mu^+ \mu^-$, which are most constraining. This is done with
the code \texttt{SuperIso}~\cite{Mahmoudi:2008tp,Mahmoudi:2009zz}
where the model input is given by {\tt{2HDMC}}.  
It should be noted that {\tt SuperIso} does not implement the recent improved calculation
of $\br(B \to X_s \gamma)$ within the 2HDM in \citere{Misiak:2020vlo},
where a better lower bound of $\sim 800\gev$ for $m_{H^\pm}$ in the 2HDM type II is quoted
(which should also apply in type III).
However,  at present there is some open discussion on the size of the theoretical uncertainties
involved on this calculation and we have
preferred to use the older computation implemented in {\tt SuperIso}. 
We have also modified the
code as to include the Higgs-Penguin type corrections in $B_s \to
\mu^+ \mu^-$ (we use the formulas in\cite{Arnan:2017lxi}) which were
not included in the original version of  
\texttt{SuperIso}.
These corrections are indeed relevant for the present work since these
Higgs-Penguin contributions are the ones containing the potential
effects from triple Higgs couplings in $B_s \to \mu^+ \mu^-$.  


\item {\bf Specific constraints on \boldmath{$m_{12}$}}\\
In the alignment limit that we are assuming through this work,  the most
relevant restrictions to $m_{12}$ are those from the theoretical
constraints previously commented and from the experimental constraints
to BR$(h \to \gamma \gamma)$.  The other sensitive observable to
$m_{12}$ is BR$(B_s \to \mu^+ \mu^-)$ but this sensitivity is very mild
and only emerges at extremely high $\tan \beta$ values,  larger
than~50~\cite{Arco:2020ucn,Arco:2022}, which we do not consider here.
(For the results shown below there is no visible impact).
Furthermore,  the constraint on $m_{12}$ from
BR$(B_s \to \mu^+ \mu^-)$ applies exclusively in the type~II
models as it has been discussed in \citere{Arco:2020ucn}.
Another observable that could be sensitive to $m_{12}$ 
is $\br(h\to Z\gamma)$.  As we have seen in the previous section,  in the alignment limit, it depends 
only on $m_{H^\pm}$, $\tan\beta$ and $m_{12}$ as the 
$\br(h\to\gamma \gamma)$. However, the sensitivity to 
$m_{12}$ is milder in the $Z\gamma $ channel than in the
$\gamma\gamma$ channel because the experimental bound 
to the signal strength is $\mu_{Z\gamma } < 3.6$ at 95\% CL \cite{PDG2022}, 
much less precise than the one in the $\gamma \gamma$ case. 
The predictions in the 2HDM of this observable in the parameter 
space explored in this paper would only exclude the low $m_{H^\pm}$ 
region with $m_{H^\pm} < 200 \gev$, corresponding to values of
$\br(h\to Z\gamma)>5\times10^{-3}$, as can be observed in the
lower right plot in \reffi{BRtogaga}. 
It should also be noted that the $S$, $T$ and $U$ parameters considered
here at one loop are not sensitive to $m_{12}$.
However,  this sensitivity might change in a two loop computation.
According to \citere{Hessenberger:2022tcx}, this sensitivity 
(given in terms of $\lambda_5$ in that reference) is very mild and it only
appears at very low $\tan\beta$ values $<1$ (see for instance Fig.\ 3 in that reference),
which on the other hand are already
strongly restricted by flavor physics. 
Hence, we have preferred to use here
the one loop formulas implemented in the {\tt 2HDMC} to study the oblique parameters.
Overall,  we then pay attention here to the 
specific analysis of $m_{12}$ from  the two most restricting 
requirements: the theoretical ones and the experimental constraints on 
BR$(h \to \gamma \gamma)$.  To apply the latter we compare the 
prediction of  BR$(h \to \gamma \gamma)$  within the 2HDM in the
alignment limit with the experimentally measured value.
The experimental data, as averaged in \citere{PDG2022}, is given by
\begin{align}
  \mu_{\ga\ga}^{\rm exp} = \frac{\sigma(pp\to h)_{\rm exp}\ \br(h \to \ga\ga)_{\rm exp}}
                               {\sigma(pp\to h)_{\rm SM}\ \br(h \to \ga\ga)_{\rm SM}} \nonumber \\ 
                                 = 1.10 \pm 0.07\,.
\label{mugaga-exp}
\end{align}
Here $\br(h \to \ga\ga)_{\rm SM}$ denotes the ``SM limit''
as defined above.
This can readily be compared to the theory prediction
by: 
 \begin{equation}
 \mu_{\gamma \gamma}^{\rm 2HDM}= \frac{{\rm BR}(h \to \gamma \gamma)_{\rm 2HDM}}{{\rm BR}(h \to \gamma \gamma)_{\rm SM}}\,.
 \label{mugaga}
 \end{equation}
 
Notice that in the alignment limit, the ratio of the production
cross section of $h$ respect to the SM present in the definition of
the signal strength cancels and  $\mu_{\ga\ga}$ is then given just in
terms of the BRs.
\end{itemize}

The main difference of the analysis presented below with
respect to the ones in \citeres{Arco:2020ucn,Arco:2022}
is that we focus now on the alignment limit (which we
did not in our previous works) and
that the $\mu_{\gamma \gamma}$ is imposed
{\it separately} as a specific constraint on $m_{12}$ (and not only
in the sum of all LHC rate measurements).  This choice is justified,  since
we are investigating the possible restrictions on $m_{12}$,  and 
$\br(h \to \ga\ga)$ is effectively the only sensitive quantity to
$m_{12}$.  Just looking at the overall LHC rate measurement would dilute
this sensitivity.
As discussed above, we first apply constraints from
theory,  EWPO,  colliders,  and flavor observables,  from now on
called ``other constraints''.
In the final step we will apply the constraints of $\br(h \to \ga\ga)$
and conclude on the $m_{12}$ values that are allowed by these  other
constraints, but disallowed  by  $\mu_{\gamma \gamma}$.
Since these other constraints depend on the specific type of 2HDM, we
will present the complete analysis separately for each Yukawa
type. This will allow us to conclude on the possible
restrictions for the
$m_{12}$ values in the four 2HDM types. 


\section{Results}

The results of our analysis are collected in \reffis{Figtypes-tb2},
\ref{Figtypes-tb5-10-50} and \ref{Figtypes-Splitt-tb2}.  In all the
plots we show the $\bar m$-$\MHp$ plane, where $m_{12}$ is displayed
  at the top of each plot.
The four columns in each of the figures represent the results in
the four Yukawa types,  correspondingly.
The colored regions in these plots (except red, see below)
correspond to the regions disallowed at the $2 \sigma$ level by 
the ``other constraints'':
light grey regions are disallowed by the
theoretical constraints,   light blue regions are disallowed by LHC
Higgs searches,  yellow regions are disallowed by flavor
observables (the two areas light blue and yellow together give the light
green areas),  dark grey regions are disallowed by the LHC rate
measurements.  It should be noted 
that light blue,  yellow and green areas are only plotted inside the
allowed region by the theoretical constraints and that the light grey areas
in the lower right part of these plots are not clearly visible because
they are hidden below the dark grey areas.  
The areas in white (disregarding red for now) represent
the $2\sigma$ allowed regions by the previously called ``other
constraints''.
On top of these allowed  areas,  we include the predicted
allowed area by 
$\mu_{\gamma \gamma}$, which  is displayed in red to be
clearly distinguishable.  These red veils with conical shape
differentiate the $1\sigma$ allowed regions,  see \refeq{mugaga-exp},
which are limited by
dotted lines and displayed in a
darker red color, and the broader $2\sigma$
allowed regions, which are limited by dashed lines
and displayed in a lighter red color. The solid red
line is the contour line for $\mu_{\gamma \gamma}$ corresponding to the
central experimental value.  
Notice that those red cones correspond to that shown in the lower left plot of \reffi{BRtogaga}.
Following our description, the red shaded areas found on top of the
areas that are allowed by the other constraints are in agreement with
all experimental data. 
On the other hand, the areas that remain white after application of the $\mu_{\ga\ga}$ bounds marked in transparent red are the areas of
our interest here in order 
to conclude on the additional constraints on $m_{12}$ from $\mu_{\ga\ga}$.
It is precisely these ``reduced white regions" what define the final disallowed intervals in $m_{12}$ that we are
looking for.  

The results in \reffis{Figtypes-tb2}, \ref{Figtypes-tb5-10-50} and
\ref{Figtypes-Splitt-tb2} cover the three generic scenarios considered
here that take into account the three qualitative different patterns of
the BSM Higgs boson masses, which we classify and name accordingly by:  
\begin{itemize}
\item{\bf Deg} \\
This scenario sets all the BSM Higgs boson masses to be fully degenerate with:
$m_A=m_H=m_{H^\pm}$.  This is designed  to easily evade the EWPO constraints. 
\item{\bf Par Deg} \\
This scenario sets the BSM Higgs boson masses to be just partially
degenerate with $m_A=m_{H^\pm}$ but $m_H$ different.  In this scenario
we consider two different cases: 1)~$m_H$ fixed to a particular
numerical value and 2)~$m_H=\bar m$.  Again the dominant EWPO
constraints from the $T$ parameter are evaded.  
\item{\bf Split}\\
This scenario is designed to possibly accommodate a new world
average of $\MW$, including the recent CDF measurement~\cite{CDF2}.
In this scenario we consider non degenerate $m_A$,  $m_{H^\pm}$ and $m_H$.  We
set $m_A<m_{H^\pm}$,  but with mass values $m_A$ and 
$m_{H^\pm}$ split by a fixed quantity $\Delta m$,  as $m_A=m_{H^\pm}-\Delta m$. 
The other Higgs mass is set by: 1)~$m_H$ fixed to a particular
numerical value,  2)~$m_H=\bar m$ and 3) $m_H= m_A$. 
We explore different values for
the splitting between  $m_A$ and $m_{H^\pm}$ and require that the
value of $m_W$ predicted in the 2HDM to be in agreement at the
$2\sigma$  level with a possible new world average~\cite{PDG2022},
including the recent CDF measurement~\cite{CDF2}, 
  $m_W^{\rm{exp,new}} \approx 80.417 \pm 0.018 \gev$.%
\footnote{It should be noted that the values given so far in
\citere{PDG2022} are rather approximate. }
To calculate $m_W$ in the 2HDM we use the one-loop
approximation~\cite{Grimus:2008nb,Biekotter:2022abc} given in terms
of the $S$, $T$ and $U$ parameters, 
\begin{eqnarray}
m_W=&m_W^{\rm SM}\sqrt{1+\frac{\SW^2}{\CW^2-\SW^2}\Delta r'} \,\, , \nonumber \\
 \Delta r'=& \frac{\alpha}{\SW^2}
\left( -\frac{S}{2}+\CW^2 T+ \frac{\CW^2-\SW^2}{4\SW^2} U \right).
\end{eqnarray}
We take the numerical values for $\alpha$,  $m_W^{\rm SM}$,  $\CW^2 = 1 - \SW^2 = m_W^2/m_Z^2$  from \cite{PDG2022}.  
\end{itemize}

We start discussing the results in  \reffi{Figtypes-tb2}. In all
plots in this figure we set $\tb = 2$.
In the first row the plots show the results
for the scenario {\bf Deg}, in the second row for the scenario
{\bf Par Deg} with $m_H=400 \gev$ and in the third row for the scenario
{\bf Par Deg} with $m_H=\bar m$.
We first see that the patterns of the
allowed areas by the theoretical constraints (i.e.\ the complementary
areas to the light grey ones) are different in the three rows (but
identical for the four Yukawa types, as required).  In the
scenario {\bf Deg} these allowed areas appear as diagonal corridors
whose widths get narrower for larger values of $\tan\beta$ up to
getting totally closed for about $\tan\beta>10$ (not shown here).  For
more details on these corridors see \citeres{Arco:2020ucn,Arco:2022}.
The second row,  for the {\bf Par Deg} scenario with fixed $m_H=400 \gev$,
shows that the allowed regions by the theoretical constraints are
not corridors  but they display rather as windows  with
sort of arrow tip shape placed at the lower left  corner in these
planes. 
In the third row,  corresponding to the {\bf Par Deg}  case and setting
$m_H={\bar m}$,  the allowed regions by the theoretical constraints
display again as diagonal corridors,  but wider in this scenario than in
the {\bf Deg} case.  The dark grey areas are very similar in all the
plots.  The experimentally allowed/disallowed  areas differentiating the
most the four 2HDM types are the ones from flavor observables.  The
disallowed areas 
by flavor physics are very similar in type~I and~IV as well as
in type~II and~III, 
see also the discussion in
\citere{Arco:2022}.  The total allowed areas by the ``other
constraints''  are larger
in type~II and~III than in type~I and~IV.   
Overlaid are the red areas allowed by $\mu_{\ga\ga}$, which are
identical in all plots.
Consequently,  as discussed above,  the ``reduced  white areas'', on the other hand, are excluded
  at the $2\,\sigma$~level by $\mu_{\ga\ga}$.

\end{multicols}
\myfigure{
\includegraphics[width=1\columnwidth]{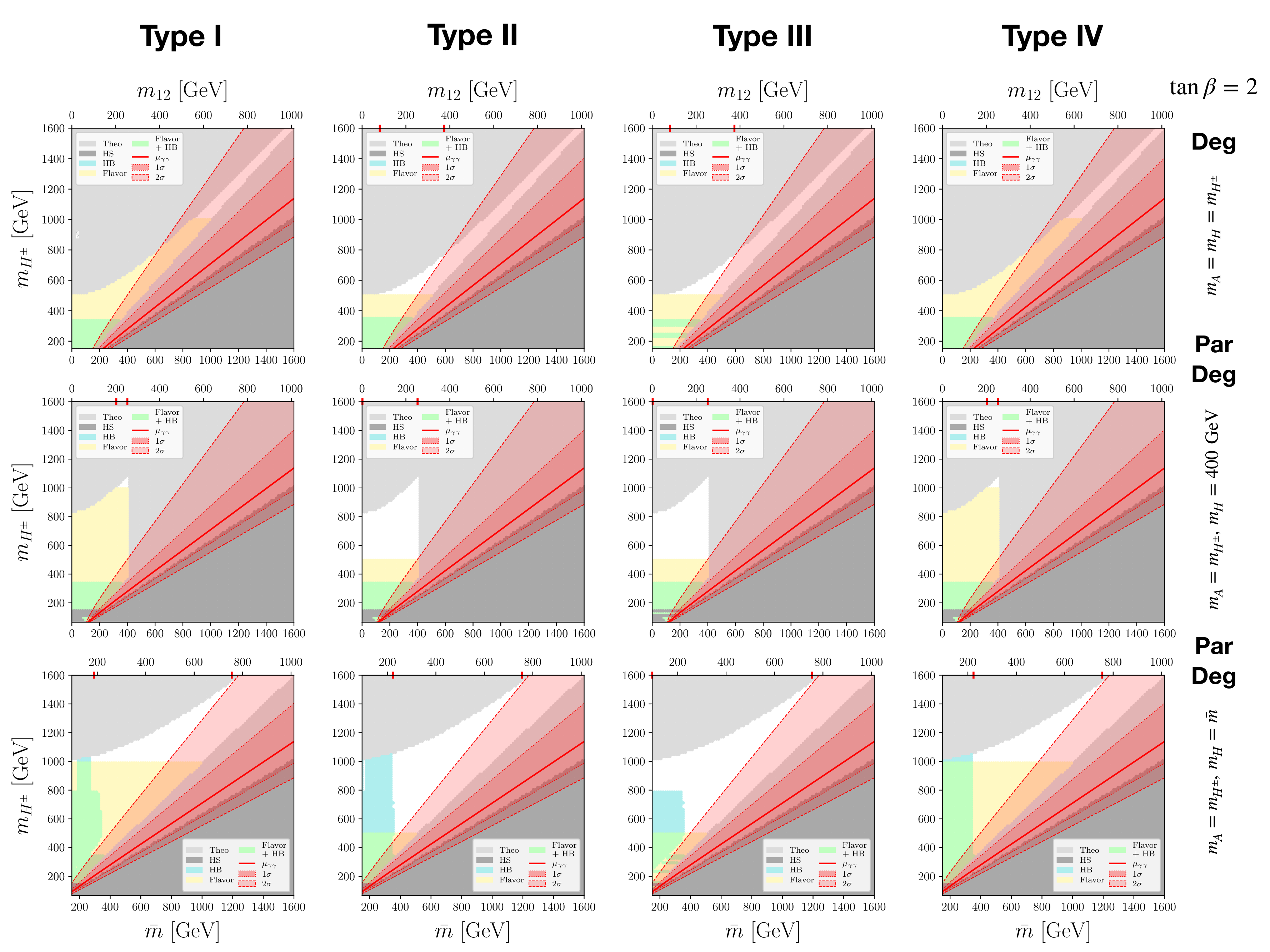}
\figcaption{Constraints in the four 2HDM  types for the {\bf Deg} and
  {\bf Par Deg} cases with  $\tan\beta=2$ (see text).}
\label{Figtypes-tb2}
}
\begin{multicols}{2}

For the purpose of the present work,  the most important conclusions
from this figure are the following: 
1)~in all cases,  all the regions
that are allowed by all  the other constraints are disallowed by
$\mu_{\gamma \gamma}$ at the $1\sigma$ level; but,  in contrast,  
2)~if we require compatibility with $\mu_{\gamma \gamma}$ at the $2\sigma$ level, 
there are some regions that are allowed by all the other constraints but
are disallowed by  $\mu_{\gamma \gamma}$
(the ``reduced white areas'') (and some other regions that are
allowed by both, the other constraints and $\mu_{\gamma \gamma}$).
The ``reduced white areas''
appear in all the studied scenarios in this figure for type~II and type~III, 
and
in the {\bf Par Deg} scenarios
for type~I and~IV.  Therefore, the final
constraints on $m_{12}$ can be extracted from the interval that is
projected on the top axis of these  plots from the `reduced white regions" and the
intercept with the red region,  that are marked in the figures with
red ticks in the upper axis.  Thus,  in the three scenarios
analyzed, for $\tb = 2$ and $\cba = 0$ we get the following disallowed
intervals in $m_{12}$  by $ \mu_{\gamma \gamma}$:   
\inter{81}{374}~GeV in type~II and~III for the {\bf Deg} case; 
\inter{202}{253}~GeV in type~I and~IV,  and \inter{0}{253}~GeV in type~II
and~III for the {\bf Par Deg} case with $m_H=400\gev$;  and 
\inter{186}{755}~GeV in type~I,  \inter{223}{755}~GeV in type~II,
\inter{0}{755}~GeV in type~III and \inter{95}{253}~GeV in type~IV for
the {\bf Par Deg} case with $m_H=\bar m$. 

\end{multicols}
\myfigure{
\includegraphics[width=1\columnwidth]{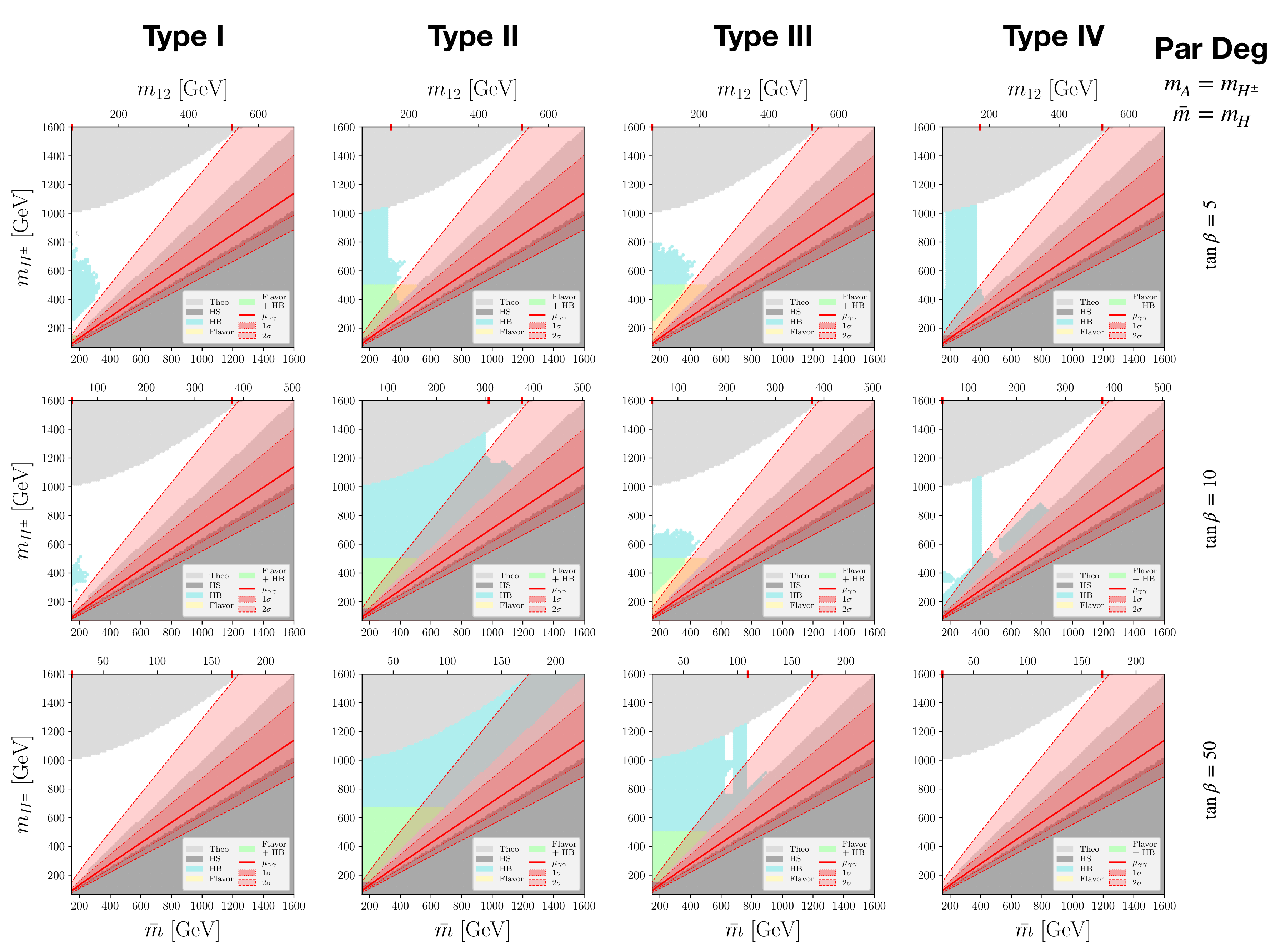}
\figcaption{Constraints in the four 2HDM  types for the  {\bf Par Deg}
  case with  $\tan\beta=5,\ 10,\ 50$ (see text). } 
\label{Figtypes-tb5-10-50}
}
\begin{multicols}{2}

Next we discuss the results of \reffi{Figtypes-tb5-10-50},
where we
explore the scenario {\bf Par Deg} with $m_H={\bar m}$ for
different values of $\tan \beta= 5, 10, 50$ in the upper, middle and
lower row, respectively.  We have not explored 
values of $\tan \beta >2$ in the other two scenarios because the
theoretical constraints effectively close
the allowed area.
The red areas, displaying the regions allowed by $\mu_{\ga\ga}$ are
by construction the same and identical to \reffi{Figtypes-tb2}.
In this
figure there are regions allowed by the ``other constraints" in all the plots (except in type~II for
$\tan\beta=50$), and consequently one can extract constraints on
$m_{12}$ from all of them. The
regions allowed by the ``other constraints'' are wider that in
the previous figure.  As before the pattern of
these regions are more
similar in the type~I and~IV  as well as in type~II and~III.
Since we have related in this figure $m_{12}$ to $m_{H}$ via the
setting $m_H={\bar m}$,  not all the derived constraints on $m_{12}$
appear to come from $\mu_{\gamma \gamma}$ but instead from  the
other constraints.  However, this implies
really limitations on $m_H$, and not on $m_{12}$.
For instance, the value $m_{12}=0$ is not included in this figure
(whereas,  it did appear in some of the plots in the previous figure) since
a very light $m_H$ is never allowed by the collider constraints
and, furthermore, by definition $m_H>m_h\sim125\gev$.   In
summary,  for the scenario {\bf Par Deg} and $\MH = \bar m$
we get the following disallowed intervals in $m_{12}$ by
$\mu_{\gamma \gamma}$:   
\inter{66}{523}~GeV in type~I and type~III,
\inter{148}{523}~GeV in type~II and \inter{174}{523}~GeV in type~IV
for $\tan\beta=5$;
\inter{47}{376}~GeV in type~I,~III and~IV and \inter{307}{376}~GeV in
type~II for $\tan\beta=10$; and 
\inter{21}{169}~GeV in type~I and~IV, \inter{109}{169}~GeV in type~III
for $\tan\beta=50$. 

\end{multicols}
\myfigure{
\includegraphics[width=1\columnwidth]{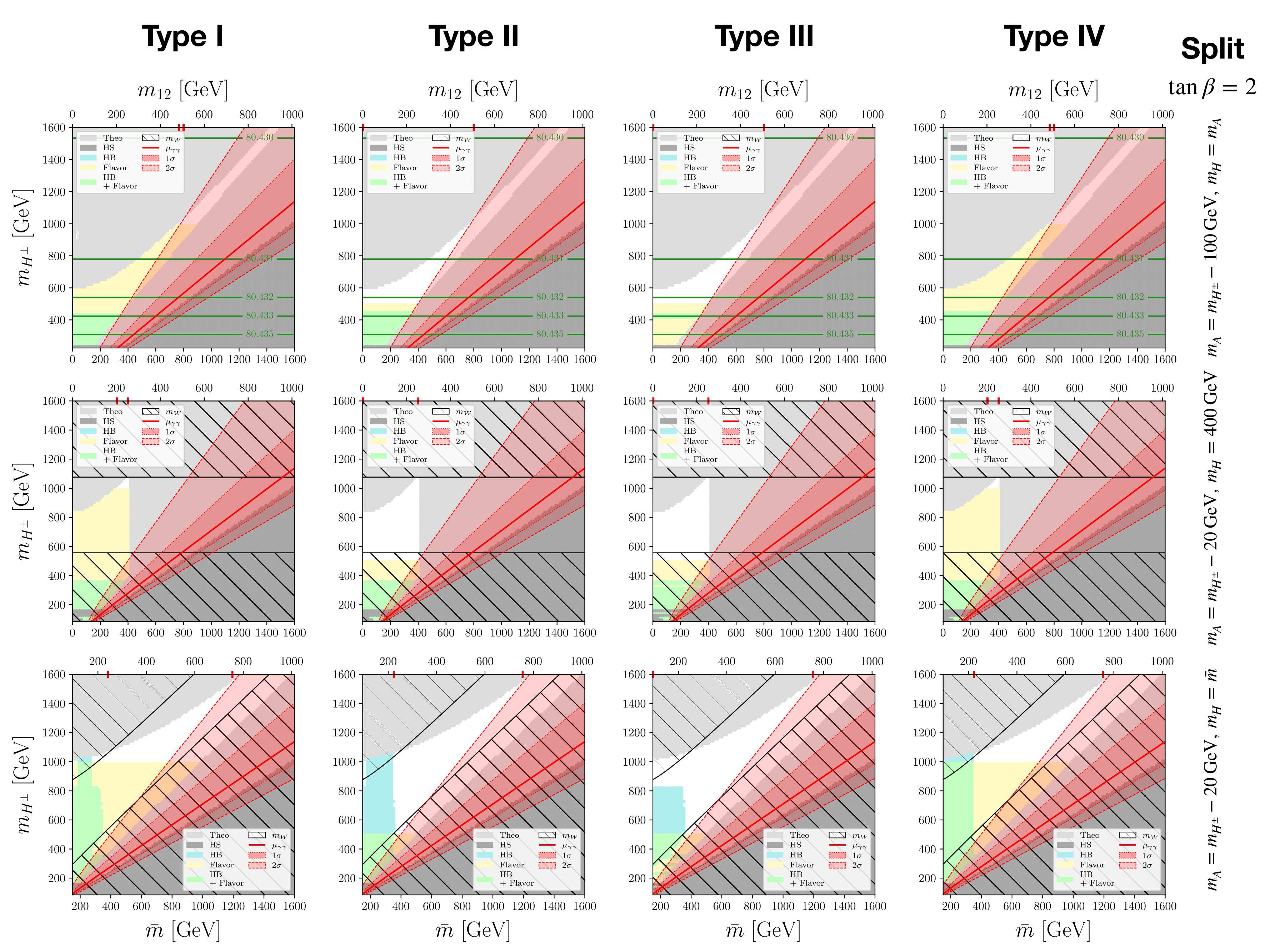}
\figcaption{Constraints in the four 2HDM  types for the {\bf Split}
  case with  $\tan\beta=2$ (see text).  }
\label{Figtypes-Splitt-tb2}
}
\begin{multicols}{2} 
Finally, we discuss the results in  \reffi{Figtypes-Splitt-tb2}. These
explore the scenario with split  
$m_{H^\pm}$ and $m_A$ values,  such that  $m_A=m_{H^\pm} - \Delta m$.  We
have set $\tan \beta=2$ in this figure and
present three cases in which we find
agreement with a possible new world average $m_W^{\rm exp,new}$ at the
$2\sigma$ level.%
\footnote{
Analyses in the 2HDM taking into account the recent CDF measurement of
$m_W$ can be found, e.g., in \citeres{Song:2022xts,Bahl:2022xzi,Babu:2022pdn}.}
~The disallowed areas by  $m_W$ are shown in these
plots by the striped regions,  which by construction are identical
in all four Yukawa types.  In the first row we set
$\Delta m=100 \gev$ and $m_H=m_A$,  in the second row $\Delta m=20 \gev$
and $m_H=400 \gev$,  and in the third row
$\Delta m=20 \gev$ and $m_H={\bar m}$.  We see in the first
row that $m_W^{\rm exp,new}$ yields no
additional constraints, i.e.\ no striped region appears.
In fact, the specified horizontal contour lines with the
predicted $m_W$ in the 2HDM are all compatible with $m_W^{\rm exp,new}$
within $2\sigma$.  Therefore, the conclusions on the $m_{12}$
constraints are very similar to  the fully degenerate case of
\reffi{Figtypes-tb2}.  Other choices for $\Delta m$ in the interval
\inter{60}{110}~GeV with $\MH = \MA$ yield
a similar result for the regions allowed by the ``other constraints".
$\Delta m$ values outside this interval do not exhibit
allowed regions by  $m_W^{\rm exp,new}$ and
result in a fully striped plots.
In the second row we have chosen $\Delta m = 20 \gev$ and
$\MH = 400 \gev$. In this case the parameter space allowed by
$m_W^{\rm exp,new}$ is a horizontal band from $\MHp = 572 \gev$ to
$\MHp = 1062 \gev$. 
Since this band contains
almost all the region allowed
by all the other constraints,
the conclusions on the final constraints on $m_{12}$ are 
very similar to results found in the second row of
\reffi{Figtypes-tb2}.
In the third row we set $\Delta m = 20 \gev$ and $\MH = \bar m$.
This results in a diagonal band allowed by $m_W^{\rm exp,new}$, going from
low values of $\bar m$, $\MHp$ to large values of these parameters.
The intersection between this diagonal band
and the ``reduced white area" provides the joint allowed region.  The projection of
this final allowed region on to the upper axis and the intercept with
the red region provides the wanted constraints on $m_{12}$.  This gives
the following disallowed intervals:
\inter{486}{506}~GeV in type~I and~IV and \inter{0}{506}~GeV in type~II
and~III for $\Delta m=100\gev$ and $m_H=m_A$; 
\inter{202}{253}~GeV in type~I and~IV and \inter{0}{253}~GeV in type~II
and~III for $\Delta m=20\gev$ and $m_H=400\gev$; and 
\inter{242}{755}~GeV in type~I, \inter{223}{755}~GeV in type~II and~IV
and \inter{94}{755}~GeV in type~III for $\Delta m=20\gev$ and $m_H=\bar m$.


\section{Conclusions}

In the present paper
we have analyzed possible constraints on the $Z_2$ soft-breaking
parameter $m_{12}$ in the 2HDM, where we have focused on the alignment limit.
we have demonstrated that it is possible to
obtain constraints on $m_{12}$
from the comparison of the 
theoretical prediction of BR$(h \to \gamma \gamma)$
with the present data on the di-photon signal strength.
The results of these constraints are summarized  
in \reffis{Figtypes-tb2}, \ref{Figtypes-tb5-10-50} and
\ref{Figtypes-Splitt-tb2}.  The new intervals of $m_{12}$ that are
disallowed at $2\sigma$  by $\mu_{\gamma \gamma}$ while being allowed by
all the other constraints have been derived for the four 2HDM types, in
several scenarios for the heavy Higgs boson masses,  as summarized in
the previous section. 
In particular we find that taking into account the new measurement
  of $m_W^{\rm CDF}$ has only a mild impact on the derived limits on $m_{12}$.

\subsection*{Acknowledgements}
\begingroup 
The present work has received financial support:
 from  the grant IFT Centro de Excelencia Severo Ochoa CEX2020-001007-S
  funded by MCIN/AEI/10.13039/501100011033;
from  the
grant PID2019-110058GB-C21 funded by
MCIN/AEI/10.13039/501100011033 and by "ERDF A way of making Europe"; from the Spanish
``Agencia Estatal de Investigaci\'on'' (AEI) and the EU ``Fondo Europeo de
Desarrollo Regional'' (FEDER) 
through the project PID2019-108892RB-I00/AEI/10.13039/501100011033;
from the European Union's Horizon 2020 research and innovation
programme under the Marie Sklodowska-Curie grant agreement No 674896 and
No 860881-HIDDeN; and from the FPU grant with code FPU18/06634. 
\endgroup

\bibliographystyle{unsrt}

\end{multicols}

\end{document}

%% file: paperdef.tex
\newcommand{\lsim}
{\;\raisebox{-.3em}{$\stackrel{\displaystyle <}{\sim}$}\;}

\newcommand\al{\alpha}
\newcommand\be{\beta}
\newcommand\ga{\gamma}
\newcommand\tb{\tan\beta}

\newcommand\cba{c_{\beta - \alpha}}

\newcommand\CBA{\cos(\be-\al)}

\newcommand\ReDiag{\mathop{%
  \raise .5pt\hbox{[}%
  \widetilde{\mathrm{Re}}%
  \raise .5pt\hbox{]}}}
\newcommand\ReOffDiag{\mathop{%
  \raise .5pt\hbox{$\llbracket$}%
  \widetilde{\mathrm{Re}}%
  \raise .5pt\hbox{$\rrbracket$}}}

\newcommand\SW{s_\mathrm{w}}
\newcommand\CW{c_\mathrm{w}}
\newcommand\MW{m_W}

\newcommand\MH{m_H}
\newcommand\MA{m_A}
\newcommand\MHp{m_{H^\pm}}

\newcommand\msq{m_{12}^{2}}

\newcommand\refeq[1]{Eq.~(\ref{#1})}

\newcommand\refta[1]{Tab.~\ref{#1}}

\newcommand\citere[1]{Ref.~\cite{#1}}
\newcommand\citeres[1]{Refs.~\cite{#1}}

\newcommand{\CP}{{\cal CP}}
\newcommand{\cp}{{\CP}}

\newcommand{\gev}{\,\, \mathrm{GeV}}

\newcommand\HB{\texttt{HiggsBounds}}

\newcommand{\br}{\text{BR}}
\newcommand{\De}{\Delta}

\newcommand{\sig}{\sigma}

\def\reffi#1{\mbox{Fig.~\ref{#1}}}
\def\reffis#1{\mbox{Figs.~\ref{#1}}}

\def\ga{\gamma}

\def\la{\lambda}

\newcommand{\lahHpHm}{\ensuremath{\la_{hH^+H^-}}}

\newcommand{\inter}[2]{\ensuremath{[#1, #2]}}

\definecolor{Orange}{rgb}{1.0,0.40,0.}
\definecolor{Lightblue}{cmyk}{0.9,0.1,0.1,0.3}
\definecolor{dgelborange}{cmyk}{0.,0.3,0.5, 0.}
\definecolor{Lila}{rgb}{0.5,0.,1}
\definecolor{Darkgreen}{rgb}{0.,.7,0.2}

%% file: Letter-m12-221121.bbl
\begin{thebibliography}{99} 

\bibitem{Gunion:1989we}
J.~F.~Gunion, H.~E.~Haber, G.~L.~Kane and S.~Dawson,
Front. Phys. \textbf{80} (2000), 1-404,
SCIPP-89/13. Erratum: [arXiv:hep-ph/9302272 [hep-ph]].


\bibitem{Aoki:2009ha} 
  M.~Aoki, S.~Kanemura, K.~Tsumura and K.~Yagyu,
  Phys.\ Rev.\ D {\bf 80}, 015017 (2009)
  [arXiv:0902.4665 [hep-ph]].


\bibitem{Branco:2011iw}
  G.~C.~Branco, P.~M.~Ferreira, L.~Lavoura, M.~N.~Rebelo, M.~Sher and
  J.~P.~Silva, 
  Phys.\ Rept.\  {\bf 516} (2012) 1
  [arXiv:1106.0034 [hep-ph]].
  

\bibitem{PDG2022}
R.~L.~Workman \textit{et al.} [Particle Data Group],
PTEP \textbf{2022} (2022), 083C01


\bibitem{Glashow:1976nt} 
  S.~L.~Glashow and S.~Weinberg,
  Phys.\ Rev.\ D {\bf 15}, 1958 (1977).
  

\bibitem{Arco:2020ucn}
F.~Arco, S.~Heinemeyer and M.~J.~Herrero,
Eur. Phys. J. C \textbf{80} (2020) no.9, 884
[arXiv:2005.10576 [hep-ph]].

\bibitem{Arco:2022}
F.~Arco, S.~Heinemeyer and M.~J.~Herrero,
Eur. Phys. J. C \textbf{82}, no.6, 536 (2022)
[arXiv:2203.12684 [hep-ph]].


\bibitem{Arco:2021bvf}
F.~Arco, S.~Heinemeyer and M.~J.~Herrero,
Eur. Phys. J. C \textbf{81}, no.10, 913 (2021)
[arXiv:2106.11105 [hep-ph]].

\bibitem{Ellis:1975ap}
J.~R.~Ellis, M.~K.~Gaillard and D.~V.~Nanopoulos,
Nucl. Phys. B \textbf{106}, 292 (1976)

\bibitem{Shifman:1979eb}
M.~A.~Shifman, A.~I.~Vainshtein, M.~B.~Voloshin and V.~I.~Zakharov,
Sov. J. Nucl. Phys. \textbf{30} (1979), 711-716
ITEP-42-1979.

\bibitem{Bernon:2015qea}
J.~Bernon, J.~F.~Gunion, H.~E.~Haber, Y.~Jiang and S.~Kraml,
Phys. Rev. D \textbf{92} (2015) no.7, 075004
[arXiv:1507.00933 [hep-ph]].

\bibitem{Eriksson:2009ws}
  D.~Eriksson, J.~Rathsman and O.~St{\aa}l,
  Comput.\ Phys.\ Commun.\  {\bf 181} (2010) 189
  [arXiv:0902.0851 [hep-ph]].

\bibitem{Arhrib:2003vip}
A.~Arhrib, M.~Capdequi Peyranere, W.~Hollik and S.~Penaranda,
Phys. Lett. B \textbf{579} (2004), 361-370
[arXiv:hep-ph/0307391 [hep-ph]].

\bibitem{Peskin:1990zt}
  M.~E.~Peskin and T.~Takeuchi,
  Phys.\ Rev.\ Lett.\  {\bf 65} (1990) 964.

\bibitem{Peskin:1991sw}
  M.~E.~Peskin and T.~Takeuchi,
  Phys.\ Rev.\ D {\bf 46} (1992) 381.
  
\bibitem{Grimus:2007if}
W.~Grimus, L.~Lavoura, O.~M.~Ogreid and P.~Osland,
J. Phys. G \textbf{35} (2008), 075001
[arXiv:0711.4022 [hep-ph]].

\bibitem{Grimus:2008nb}
W.~Grimus, L.~Lavoura, O.~M.~Ogreid and P.~Osland,
Nucl. Phys. B \textbf{801} (2008), 81-96
[arXiv:0802.4353 [hep-ph]].

\bibitem{CDF2}  
T.~Aaltonen \textit{et al.} [CDF],
Science \textbf{376} (2022) no.6589, 170-176.
  
\bibitem{Bhattacharyya:2015nca}
  G.~Bhattacharyya and D.~Das,
  Pramana {\bf 87} (2016) no.3,  40
  [arXiv:1507.06424 [hep-ph]].

 \bibitem{Bechtle:2008jh}
  P.~Bechtle, O.~Brein, S.~Heinemeyer, G.~Weiglein and K.~E.~Williams,
  Comput.\ Phys.\ Commun.\  {\bf 181} (2010) 138
  [arXiv:0811.4169 [hep-ph]].

\bibitem{Bechtle:2020pkv}
P.~Bechtle, D.~Dercks, S.~Heinemeyer, T.~Klingl, T.~Stefaniak, G.~Weiglein and J.~Wittbrodt,
[arXiv:2006.06007 [hep-ph]].

\bibitem{Bechtle:2013xfa}
  P.~Bechtle, S.~Heinemeyer, O.~Stål, T.~Stefaniak and G.~Weiglein,
  Eur.\ Phys.\ J.\ C {\bf 74} (2014) no.2,  2711
  [arXiv:1305.1933 [hep-ph]].

\bibitem{Bechtle:2020uwn}
P.~Bechtle, S.~Heinemeyer, T.~Klingl, T.~Stefaniak, G.~Weiglein and
J.~Wittbrodt, 
Eur. Phys. J. C \textbf{81} (2021) no.2, 145
[arXiv:2012.09197 [hep-ph]].

\bibitem{Enomoto:2015wbn}
  T.~Enomoto and R.~Watanabe,
  JHEP {\bf 1605} (2016) 00.2
  [arXiv:1511.05066 [hep-ph]].

\bibitem{Arbey:2017gmh}
  A.~Arbey, F.~Mahmoudi, O.~St{\aa}l and T.~Stefaniak,
  Eur.\ Phys.\ J.\ C {\bf 78} (2018) no.3,  182.
  [arXiv:1706.07414 [hep-ph]].
  
\bibitem{Abe:2015oca}
T.~Abe, R.~Sato and K.~Yagyu,
JHEP \textbf{07} (2015), 064
[arXiv:1504.07059 [hep-ph]].

\bibitem{Mahmoudi:2008tp}
  F.~Mahmoudi,
  Comput.\ Phys.\ Commun.\  {\bf 180} (2009) 1579.
  [arXiv:0808.3144 [hep-ph]].

\bibitem{Mahmoudi:2009zz}
F.~Mahmoudi,
Comput. Phys. Commun. \textbf{180} (2009), 1718-1719.

\bibitem{Misiak:2020vlo}
M.~Misiak, A.~Rehman and M.~Steinhauser,
JHEP \textbf{06} (2020), 175
[arXiv:2002.01548 [hep-ph]].

\bibitem{Arnan:2017lxi}
  P.~Arnan, D.~Becirevic, F.~Mescia and O.~Sumensari,
  Eur.\ Phys.\ J.\ C {\bf 77} (2017) no.11,  796.
  [arXiv:1703.03426 [hep-ph]].
  
\bibitem{Hessenberger:2022tcx}
S.~Hessenberger and W.~Hollik,
[arXiv:2207.03845 [hep-ph]].

\bibitem{Biekotter:2022abc}
T.~Biek\"otter, S.~Heinemeyer and G.~Weiglein,
[arXiv:2204.05975 [hep-ph]].

\bibitem{Song:2022xts}
H.~Song, W.~Su and M.~Zhang,
[arXiv:2204.05085 [hep-ph]].

\bibitem{Bahl:2022xzi}
H.~Bahl, J.~Braathen and G.~Weiglein,
[arXiv:2204.05269 [hep-ph]].

\bibitem{Babu:2022pdn}
K.~S.~Babu, S.~Jana and Vishnu~P.~K.,
[arXiv:2204.05303 [hep-ph]].
  
\end{thebibliography}
